\documentclass[draft]{agujournal2019}
\usepackage{url} 
\usepackage{lineno}
\usepackage[inline]{trackchanges} 
\usepackage{soul}

\draftfalse

\journalname{Arxiv.org}

\begin{document}

%
%


\title{Evidence of potential thermospheric overcooling during the May 2024 geomagnetic superstorm}

%
%




\authors{Alok Kumar Ranjan\affil{1,2}, Dayakrishna Nailwal\affil{1}, M. V. Sunil Krishna\affil{1,3}, Akash Kumar\affil{1},  Sumanta Sarkhel\affil{1,3}}


\affiliation{1}{Department of Physics, Indian Institute of Technology Roorkee, Roorkee-247667, Uttarakhand, India}
\affiliation{2}{Space and Atmospheric Sciences Division, Physical Research Laboratory, Ahmedabad, India}
\affiliation{3}{Centre for Space Science and Technology, Indian Institute of Technology Roorkee, Roorkee-247667,
India}




\correspondingauthor{M. V. Sunil Krishna}{mv.sunilkrishna@ph.iitr.ac.in}



\begin{keypoints}
\item An all time high thermospheric NO radiative cooling flux is observed during the recent May 2024 geomagnetic superstorm.
\item A potential post-storm thermospheric overcooling is observed by TIMED/SABER, Swarm-A, -B, and -C during this superstorm. 
\item A comparison of thermospheric NO radiative cooling is also presented between the famous Halloween storms of October 2003, and May 2024 geomagnetic superstorm. 
\end{keypoints}

%
%

%
%


\begin{abstract}
[During intense geomagnetic storms, the rapid and significant production of nitric oxide (NO) followed by its associated infrared radiative emission in lower thermosphere contributes crucially to the energetics of the upper atmosphere. This makes NO infrared radiative cooling a very important phenomenon which needs to be considered for accurate density forecasting in thermosphere. This study reports the investigation of variations in thermospheric density, and NO radiative cooling during the recent geomagnetic superstorm of May 2024. A very rare post-storm thermospheric density depletion of about -23\% on May 12 was observed by Swarm-C in northern hemisphere in comparison to the prestorm condition on May 9. \textcolor{black}{This overcooling was observed despite the continuous enhancement in solar EUV (24-36 nm) flux throughout the event.} The thermospheric NO infrared radiative emission in the recovery phase of the storm is likely to be the plausible cause for this observed post-storm density depletion. Our analysis also suggests an all time high thermospheric NO radiative cooling flux up to 11.84 ergs/cm$^2$/sec during May 2024 geomagnetic superstorm, which has also been compared with famous Halloween storms of October 2003.]

\end{abstract}

\section*{Plain Language Summary}
The changing heat budget of upper atmosphere due to enhanced Joule heating, energetic particle precipitations, and NO infrared radiative cooling during geomagnetic storms may also modulate other atmospheric and ionospheric parameters that indirectly affect human life at the surface. Some examples are, thermospheric density and associated Lower Earth Orbit (LEO) satellite drags (which are particularly used for geophysical imaging), GPS navigation, exposure of thermospheric particles near ISS (International Space stations), etc. All these aspects need to be considered for a better space weather future. In this study, the variation in thermospheric density, and thermospheric NO radiative cooling have been investigated during the recent geomagnetic superstorm of May 2024. This study infers that the very enhanced NO radiative cooling in the recovery phase of this storm may have also contributed to the thermospheric density depletion in the post-storm periods. The NO radiative cooling during this storm has also been compared with Halloween storms of October 2003.

%
%

%


%
%
%
%

\section{Introduction}
\justifying
The Sun's incoming dynamic energy determines density, temperature, and structure of the Earth's upper atmosphere. Although the primary energy source of thermosphere is high energetic solar radiation (EUV and X-rays), solar wind-magnetospheric energy input predominates as a major energy source for the thermosphere during geomagnetic storms \cite{knipp2004direct}. The interaction of the solar wind and magnetosphere causes a small fraction of the solar wind's kinetic energy to be transferred and absorbed into the magnetosphere in the form of electrical energy. Evidently, a considerable portion of this absorbed energy is then released in the polar upper atmosphere or thermosphere of both the hemispheres as heating resources by currents (Joule heating) and precipitating energetic particles \cite{mayr1978some,prolss1980magnetic,richmond2021joule}. Fortunately, carbon dioxide (CO$_2$) and nitric oxide (NO) emissions, as well as mesospheric conduction in the mesosphere and lower thermosphere (MLT) contribute to cooling in Earth's upper atmosphere. They assist the thermosphere in restoring its equilibrium immediately following the perturbations caused by these heating resources \cite{roble1987global,kockarts1980nitric,mlynczak2003natural,mlynczak2005energy,mlynczak2010observations,knipp2013thermospheric,oliveira2019satellite}.

\justifying
During intense space weather events, the Earth's polar lower thermosphere experiences increased energetic particle precipitation, which produces nitric oxide (NO) \cite{barth2003global,duff2003rate,siskind2004middle}. Additionally, an increase in NO density also occurs in the lower thermosphere at low-mid latitudes due to the combined effects of the storm's Joule heating, meridional wind, and energetic solar radiation (soft X-rays; 2-7 nm) \cite{barth2009joule}. Nitric oxide (NO) is a major radiative cooling source that has a large effect on the thermospheric temperature structure. Vibrational excitation of the molecule (NO ($v$ = 0 $\rightarrow$ $v$=1 \textit{or} 2)) by collisions or chemical reactions with other atoms and molecules (O, O$_2$ and N$_2$) is the first step in the thermosphere's energy loss process \cite{hwang2003vibrational,mlynczak2021spectroscopy}. One or more infrared photons are then emitted, returning the molecule to its ground state (NO ($v$=0)). A net equilibrium is thus reached when the extra thermal kinetic energy deposited in the thermosphere is transformed into radiative energy and dissipated to space. The spatial and temporal variation in the NO radiative cooling can be distinct in different geomagnetic conditions \cite{verkhoglyadova2011ionospheric,bharti2018storm,li2019understanding,ranjan2023aspects,bag2024thermospheric}. These variations are controlled mostly by storm induced perturbations in composition (NO, and O) and temperature in thermosphere \cite{ranjan2023no}. Additionally, the behavior of the NO radiative cooling profile in the MLT regions can also be influenced during geomagnetically quiet periods by the lower atmospheric forcing (atmospheric tides, and sudden stratospheric warming events) \cite{oberheide2013impact,nischal2019solar,kumar2024influence}.

\justifying
During geomagnetic storms, there is a significant fluctuation in the composition (O/N$_2$) and density of the thermosphere on global scale \textcolor{black}{\cite{prolss2011density,zhang2014storm}.} Increased heating of the polar thermosphere causes the O/N$_2$ ratio to decrease and the thermospheric density to rise dramatically \cite{forbes1996magnetic,fuller1994response,lei2010wind}. Simultaneously, NO thermospheric radiative cooling contributes significantly to the recovery of thermospheric density to the pre-storm level by converting the additional kinetic energy of thermosphere into radiative energy, which escapes to space. Considering that NO created during geomagnetic storms can persist up to 24 hours \textcolor{black}{\cite{maeda1992heat,solomon1999auroral}}, the elevated post-storm NO thermospheric radiative cooling might potentially result in thermospheric ``overcooling", which is one of major drivers for the post-storm thermospheric density to sometimes also be lower than the pre-storm values \cite{lei2012overcooling,chen2018numerical}. 

\justifying
The enhanced neutral density during geomagnetic storm increases air drag on low Earth orbit (LEO) satellites operating in the thermosphere. To determine the satellite orbit and prevent collisions, a thorough understanding of the storm-time thermospheric neutral density response to both the heating and NO radiative cooling in thermosphere is essential. The recent geomagnetic superstorm (Dst-index $<$ -400 nT) of 10-12 May is the strongest geomagnetic storm (extreme G5) since the Halloween storms. Ground- and space-based solar observatories recorded multiple X-class solar flares and Earth-bound coronal mass ejections (CMEs) between May 10–12, 2024. \textcolor{black}{The overcooling of thermosphere is a very rare event which requires large amount of post-storm NO radiative cooling, and to the best our knowledge it has been only reported once during the famous Halloween storm of 23$^{rd}$ solar cycle (29-31 October, 2003) \cite{lei2012overcooling,chen2018numerical}. The isolated May 2024 geomagnetic superstorm has nearly same main phase duration and even more strength than the multi-phased Halloween storms of October 2003. Which makes this event very appealing for the further confirmation of overcooling effect.} This study represents a potential post-storm overcooling of thermosphere in the northern hemisphere during the recent geomagnetic superstorm of 10-12 May, 2024. Section 2 details the data sources and processing methods used in this investigation. Section 3 contains the findings and discussion of our investigation, followed by Section 4 which summarizes this study.

\section{Data Resources}
\justifying
Swarm is the first Earth observation (EO) constellation mission of the European Space Agency (ESA). The mission consists of three identical satellites called Swarm-A, Swarm-B, and Swarm-C. They were launched into a near-polar orbit on November 22, 2013. Swarm aims to observe thermospheric density and horizontal winds in addition to measuring the Earth's global geomagnetic field with unprecedented spatial and temporal resolution and precision. While Swarm-B is traveling at a higher orbit of 511 km (starting altitude) and 87.75$^\circ$ inclination angle, Swarm-A and Swarm-C constitute the lower pair of satellites flying side-by-side (1.4$^\circ$ separation in longitude at the equator) at an altitude of 462 km (starting altitude) and at an 87.35$^\circ$ inclination angle. Accelerometer sensors on Swarm-A, Swarm-B, and Swarm-C monitor non-gravitational forces. After appropriate treatments, these observations can be utilized to estimate thermospheric density data obtained using precise orbit determination (POD) \cite{visser2013thermospheric,iorfida2023swarm}. Swarm-A, -B, -C 30 seconds time resolution datasets derived from GPS accelerations are obtained from \url{ftp://thermosphere.tudelft.nl/} \cite{siemes2023new}, and are utilized to probe the thermospheric density (kg/m$^3$) variations during the most recent geomagnetic superstorm of 10-12 May, 2024.

\justifying
To examine the radiative cooling pattern caused by NO in the lower thermosphere during this event (8-13 May 2024), TIMED/SABER (Thermosphere-Ionosphere-Mesosphere Energetics and Dynamics/Sounding of the Atmosphere using Broadband Emission Radiometry) Level-2A observational data is used \textcolor{black}{\cite{mlynczak1997energetics,russell1999overview,esplin2023sounding}. SABER, an infrared radiometer, scans the Earth's limb from 400 km to the surface, measuring infrared radiance (Watt m$^{-2}$ sr$^{-1}$) in 10 unique spectral channels. Applying an Abel transform on limb radiances in the 5.3 $\mu$m channel results in vertical profiles of infrared energy loss by NO, attributable to radiative cooling and it is expressed in ergs/cm$^3$/sec \cite{mlynczak2003natural,mlynczak2005energy,mlynczak2010observations}.} The NO infrared radiative flux (NO IRF) is derived by integrating the TIMED/SABER observed NO volume emission rates (NO\_ver\_unfilt) over the altitude range of 115-250 km. The NO IRF is then used to determine the NO daily radiated power by performing a surface integral over the globe within the latitude boundaries of SABER observation (53$^\circ$S to 83$^\circ$N) during this event. \textcolor{black}{The mean NO IRF for 10$^\circ$ latitude and 180$^\circ$ longitude bins have been multiplied with respective surface area for the estimation of mean NO radiative power in the respective region. Afterwards, the mean NO radiative power of all the bins for the whole day have been added to get the estimated daily NO radiative power within the latitude limit of SABER observation.} \textcolor{black}{To observe the variation in EUV flux, Solar Heliospheric Observatory (SOHO) Solar EUV Monitor (SEM) along with TIMED/SEE (Solar EUV Experiment) datasets ate utilized.} To characterize the geomagnetic superstorm considered in this study, solar wind parameters such as plasma pressure and IMF-Bz (north-south) component, as well as geomagnetic indices (Dst-index, AE-index, ap-index, and polar cap index for Northern hemisphere (PCN-index)) are used with 1 hour and 3 hour cadence from the OMNIWeb data set (\url{https://omniweb.gsfc.nasa.gov/}) and The International Service of Geomagnetic Indices (ISGI) (\url{https://isgi.unistra.fr/whats_isgi.php}).

\section{Results and Discussions}
\subsection{Solar and geomagnetic conditions during the storm}
\justifying
Massive solar flares and CMEs surged towards the Earth between May 7-14, pushing clouds of energetic charged particles with enhanced magnetic fields, resulting in the largest solar storm to strike the planet in the last 20 years. During this period, at least seven CMEs and several powerful solar flares blasted towards the Earth. There were nearly eight strongest X-class flares peaking on 14th May with a X8.7 intensity. This intensified solar activity was caused by AR3664, an active region with a huge cluster of sunspots. The CMEs, which were traveling faster than a thousand kilometers per seconds began to reach toward Earth on May 10. Their interaction with Earth's magnetosphere resulted in a persistent geomagnetic superstorm with a G5 classification. During this geomagnetic superstorm, magnificent auroras could be seen \textcolor{black}{in low-mid latitude regions also.} Northern India and the southern United States were among the extremely low latitude regions where auroras were observed.

\begin{figure*}[hbt!]
\includegraphics[width=5.2in,height=5.2in,trim=0.2cm 2cm 0.2cm 3cm, clip]{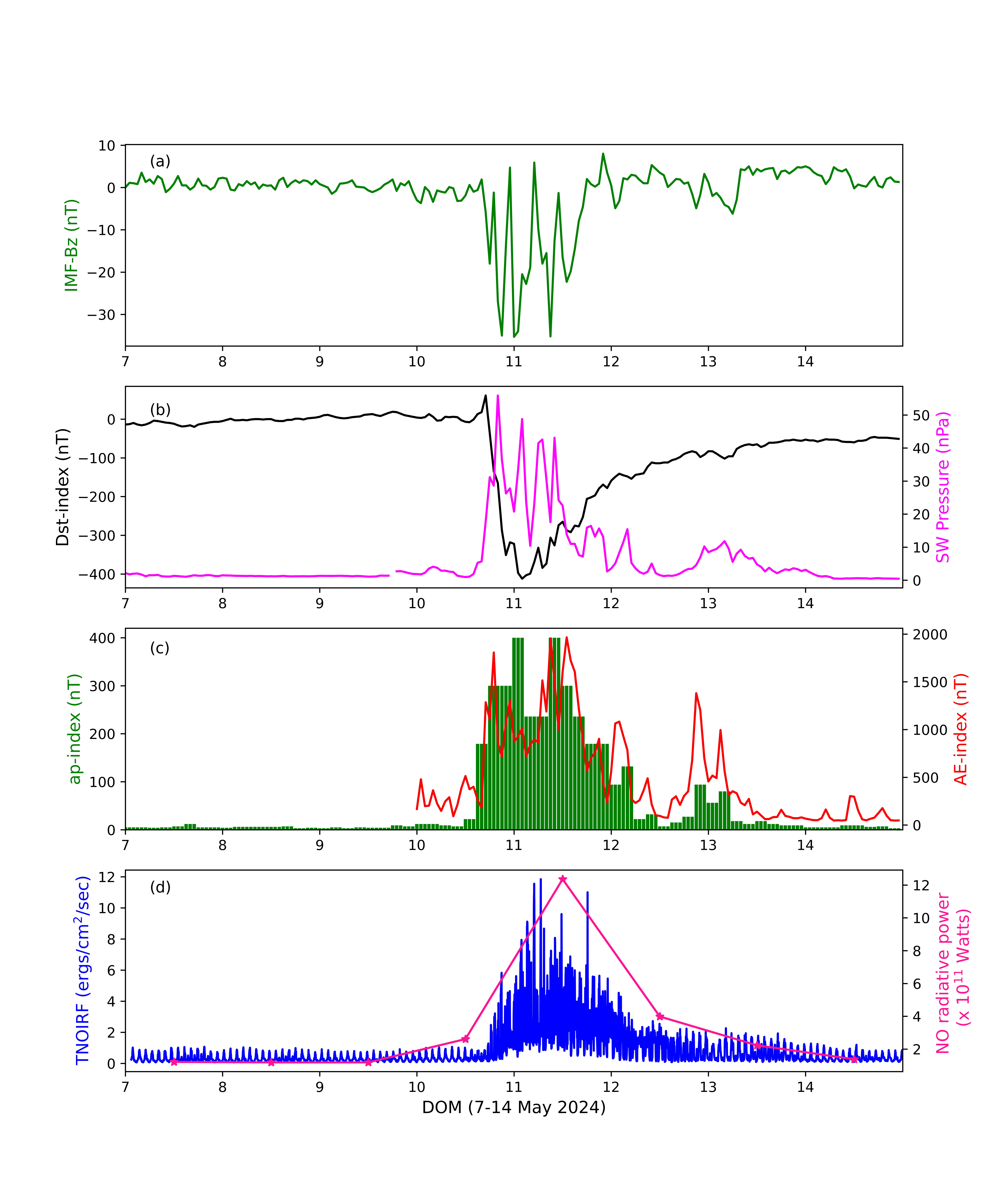}
\centering
\caption{Variations in (a) North-south component of interplanetary magnetic field (IMF-B$_z$) (green), (b) Dst-index (black) and solar wind pressure (magenta), (c) ap-index (green) and AE-index (red), and (d) TIMED/SABER observed NO infrared radiative flux or NO IRF (blue) and associated daily radiative power (magneta) throughout the considered event (7-14 May, 2024). \textcolor{black}{DOM in X-axis label represents the Date of Month.}}
\label{fig:figure1}
\end{figure*}

\justifying
Figure \ref{fig:figure1}(a-c) shows the characteristics of solar wind and geomagnetic indices throughout the duration of the superstorm (7-14 May 2024). DOM in X-axis label of this (and all the next) figure represents the Date of Month. The high pressure solar wind of about 50 hPa and southward frozen-in interplanetary magnetic field (IMF) associated with CMEs can be seen to impinge on Earth's magnetosphere. As a result, a well known positive sudden impulse (SI$^+$) \textit{or} sudden storm commencement (SSC) can be seen with a Dst-index increase of about 60 nT at 17 UT on 10$^{th}$ of May \cite{araki1994physical,gonzalez1994geomagnetic}. Shortly after, the interaction between the southward (up to -36 nT) IMF with the earth's magnetic field caused a sharp decrease in Dst-index, which represents the energization of ring current and storm's main phase. Dst-index reaches a minimum of about -412 nT at 3 UT on 11$^{th}$ of May, which is lowest Dst-index observed in the last 20 years. The substantial values of the ap (up to 400 nT) and AE (up to 2000 nT) indices demonstrate the high intensity of this geomagnetic activity on both global and local (auroral) scales. The storm's recovery phase began about 4 UT on the 11$^{th}$ of May and appears to endure for next 48 hours and more. 

\subsection{Thermospheric NO radiative cooling during the storm}
\justifying
The vast amount of energy precipitation in the polar thermosphere in the form of Joule heating and energetic particle precipitation during geomagnetic storms changes the dynamics and composition of Earth's upper atmosphere, including the enhancement of trace species like NO \cite{fuller1994response,richmond2000upper,sutton2009rapid,barth2009joule}. Simultaneously, the enhanced thermospheric kinetic energy contributes to the generation of vibrationally excited NO ($v$=0 $\rightarrow$ $v$=1) via collisions with atomic oxygen \cite{mlynczak2021spectroscopy}. This ultimately results in the emission of infrared radiation of 5.3 $\mu$m by NO ($v$=1 $\rightarrow$ $v$=0). This whole process can sometimes be accountable for emission of up to 80 \% of the joule heating energy during geomagnetic storms \cite{lu2010relationship}. Figure \ref{fig:figure1}(d) shows the variation in TIMED/SABER observed NO IRF (blue) and associated daily radiative power (between 53 $^\circ$S to 83 $^\circ$N) during the period of 7-14 May 2024. It is clearly evident in the figure that, the NO IRF starts increasing during the main phase of the storm, and it reaches about 12 ergs/cm$^2$/sec in the early UT hours of May 11. The associated enhancement in NO daily radiated power also increases up to 12.35 $\times$ 10$^{11}$ Watts on 11 May in comparison with $\sim$ 1.18 $\times$ 10$^{11}$ Watts on May 8 and 9. The daily radiative by NO exceeded 1 TW for the first time on May 11 \cite{mlynczak2024global}.

\subsection{Thermospheric density response to the storm}

\begin{figure*}[hbt!]
\includegraphics[width=5.5in,height=3.75in,trim=1.7cm 0.85cm 0cm 1.3cm, clip]{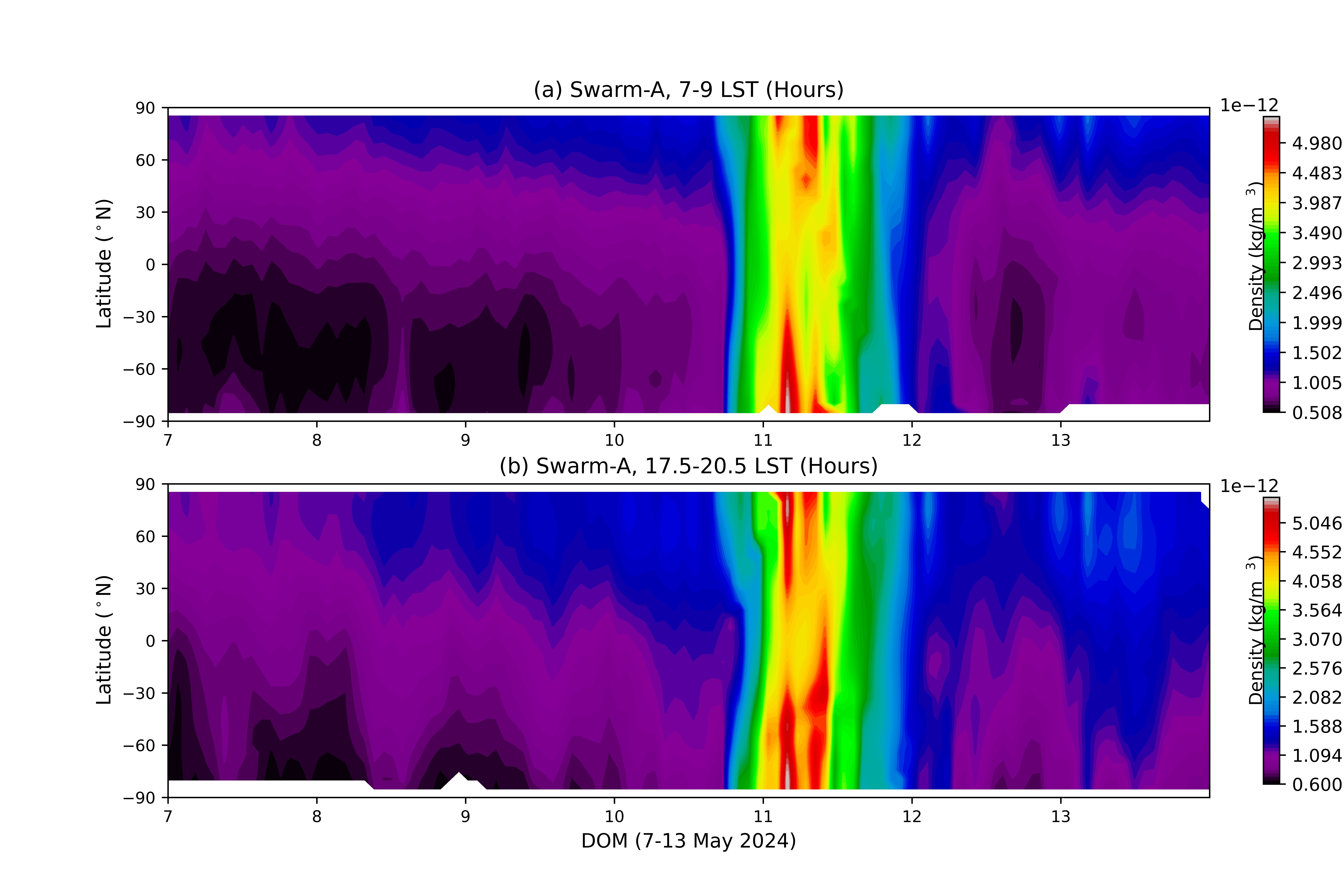}
\centering
\caption{Geodetic latitude and temporal variation in Swarm-A thermospheric density (a) in morning (6 to 9 LST), (b) evening (17.5 to 20.5 LST; 17:30 to 20:30 LST in hr:min format) from 7 to 13 May 2024.}
\label{fig:figure2}
\end{figure*}

\justifying
Joule heating during geomagnetic storms typically raises temperatures and causes upwelling or expansion of the earth's upper atmosphere. This upwelling increases the thermospheric density between 300 and 500 km above sea level. Figure \ref{fig:figure2} shows the variation in Swarm-A observed thermospheric density from 7 to 13 May. \textcolor{black}{The mass density is normalized at 490 km ($\rho$(alt) = $\rho$(z) $\times$ $\rho_{nrl}$(alt)/$\rho_{nrl}$(z); where `alt' is altitude of normalization, and z is altitude of satellite observation) by NRLMSISE-00 (Naval Research Laboratory Mass Spectrometer and Incoherent Scatter Radar Exosphere) model estimated neutral mass density \cite{picone2002nrlmsise,lei2012overcooling,van2020thermosphere}. The NRLMSISE-00 model used in this study takes the 81-day average F10.7 index (F10.7$_{81}$; centered at the day of interest), daily mean of F10.7 index (F10.7$_P$) for the previous day, and the daily average ap-index (ap) for the day of interest as the input parameters to estimate the atmospheric neutral temperature and composition including the neutral density. These input parameters have been taken for OMNIWeb data resources for each corresponding days. The altitude normalization has been done to clearly bring out the influence of polar heating during the superstorm.}



\begin{figure*}[hbt!]
\includegraphics[width=5.5in,height=2.8in,trim=0cm 0cm 0cm 0cm, clip]{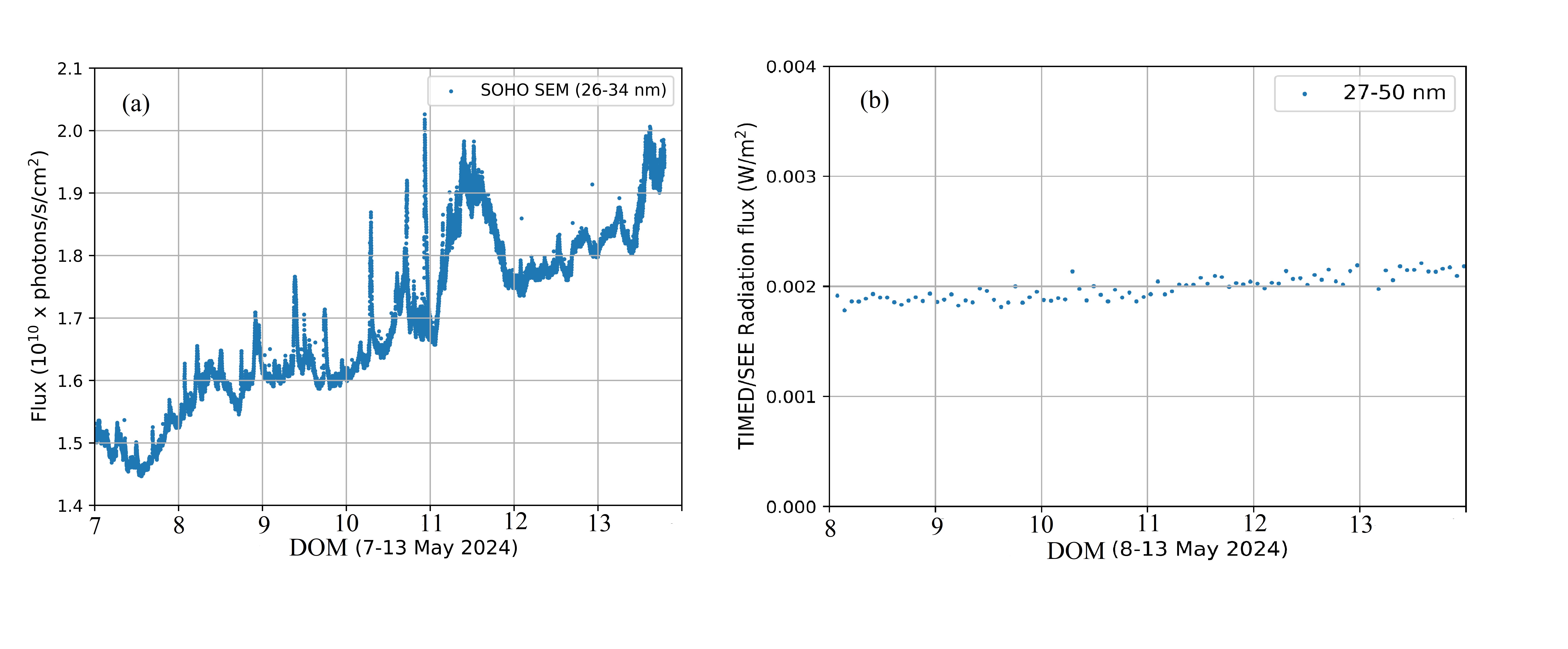}
\centering
\caption{Variation in Solar EUV fluxes as observed by (a) SOHO/SEM, and (b) TIMED/SEE throughout the event.}
\label{fig:figure3}
\end{figure*}

\justifying
\textcolor{black}{Solar EUV flux is the major heating source for the earth's upper atmosphere that can also change the thermospheric density \cite{vourlidas2018euv}. Figure \ref{fig:figure3}(a) shows the variation in the SOHO/SEM measured EUV (24-36 nm) flux from 7 to 13 of May. For simulating the thermospheric density at higher altitudes, the most effective proxy is typically the 26–34 nm integrated flux \cite{dudok2011determination}. Figure \ref{fig:figure3}(b) shows the TIMED/SEE observed radiation flux between 27-50 nm range between May 8-13. Both the figures indicate that solar EUV flux is continuously increasing throughout the event.} 

\textcolor{black}{The slow increase of thermospheric density from 7 to 10 May (before the storm commencement) in both the figures (Figure \ref{fig:figure2}(a \& b)) is likely due to the EUV flux enhancement as can be seen in Figure \ref{fig:figure3}(a \& b).} It is also evident from both the figures that, thermospheric density starts increasing in the polar regions near 17 UT on 10$^{th}$ of May in response to storm induced Joule heating, and reaches a maximum in the early UT hours of 11$^{th}$ of May. A clear latitudinal structure can also be observed in the thermospheric density enhancement during both morning (6-9) Local Solar Time (LST) and evening (17:30-20:30) LST. The density enhancement starts quicker in equatorial regions in the local early morning hours (19.2 UT on May 10; Figure \ref{fig:figure2}(a)) in comparison to evening local hours (21 UT on May 10; Figure \ref{fig:figure2}(b)), which could be due to the less ion drag effect on the storm induced meridional equatorward flow of enhanced molecular species in the early morning \cite{fuller1994response}.

\begin{figure*}[hbt!]
\includegraphics[width=5.5in,height=4.2in,trim=1.7cm 0.3cm 0.3cm 1.5cm, clip]{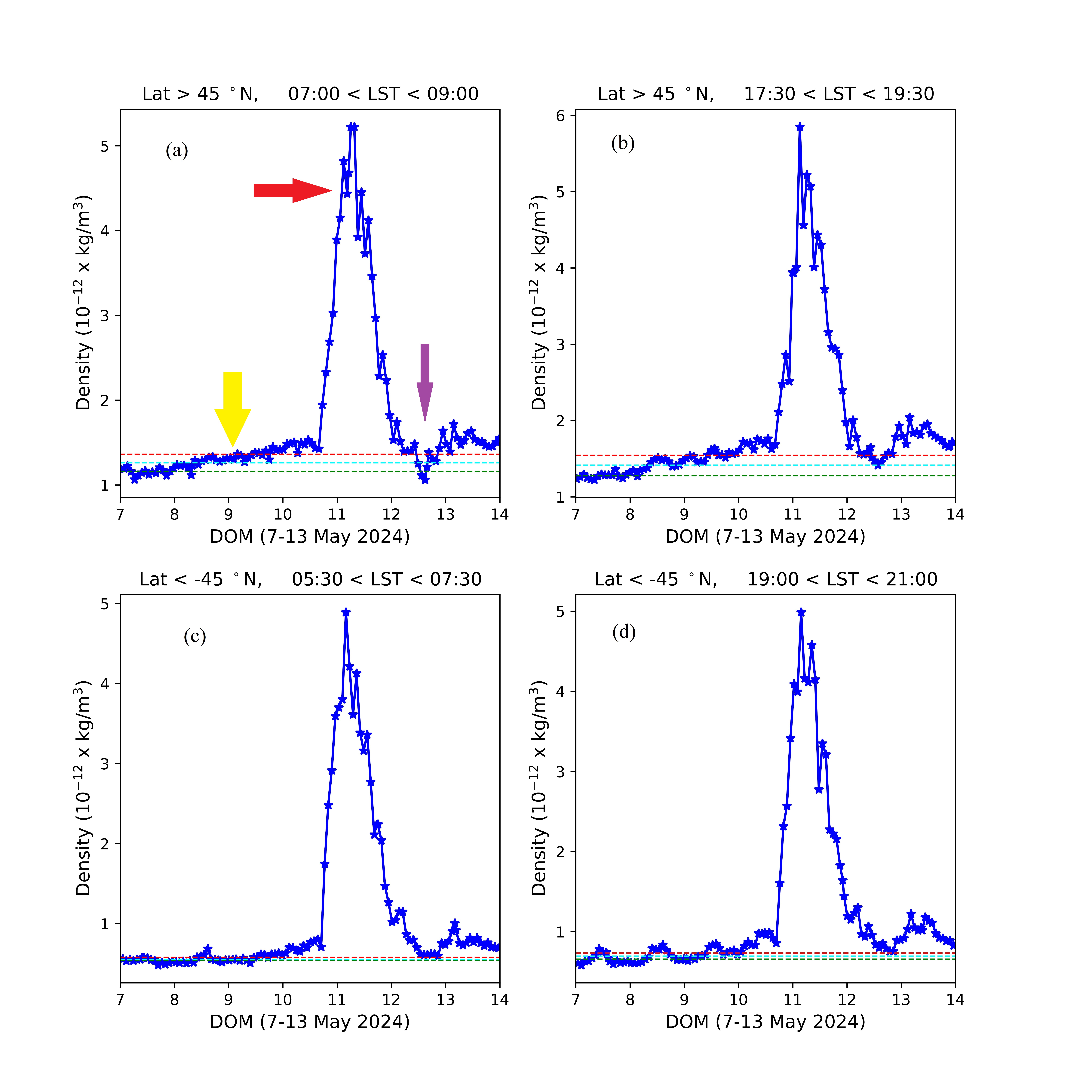}
\centering
\caption{Thermospheric density perturbations as estimated by Swarm-A (a \& b) northern polar region, and (c \& d) southern polar region, from 7-13 May 2024. The green, cyan, and red horizontal dotted lines represent the daily mean density values (within the LST and latitude limits of each figure) for 7 May, 8 May, and 9 May, respectively. (See to text for more details)}
\label{fig:figure4}
\end{figure*}

The thermospheric density also seems to be larger in the dayside, and it starts decreasing in the late UT hours of 11$^{th}$ of May, in the recovery phase of the storm. A noticeable depletion in thermospheric density can also be seen in the northern polar regions of both the figures (Figure \ref{fig:figure2}(a \& b)) at about 12 UT on 12$^{th}$ of May. The thermospheric density in northern hemisphere during this post-storm period is even lower than the pre-storm density on 9$^{th}$ of May, which is a very rare event. To make this rare post-storm density depletion more clear, \textcolor{black}{the density average line plots from Swarm-A for both the polar hemisphere considering associated LST have been shown in Figure \ref{fig:figure4}(a-d). Figure \ref{fig:figure4}(a) shows the variation in averaged density above 45 $^\circ$N and for 07:00 $<$ LST $<$ 09:00. The yellow arrow indicates the slow density enhancement caused by increasing solar EUV flux, and the red arrow shows the sudden enhancement in density during the superstorm caused by enhance Joule heating during the event. The purple coloured arrow represents the poststorm density depletion on 12$^{th}$ of May. The green, cyan, and red horizontal dotted lines represent the daily mean density values (considering the LST and latitude limits) for May 7, May 8, and May 9, respectively. Considering these green, cyan, and red horizontal lines as reference values before the superstorm \cite{zhang2019impact}, the relative thermospheric density depletion on 12$^{th}$ of May reached up to -8.46\%, -15.8\%, and -22\%, respectively in the northern polar regions. Similarly, for Figure \ref{fig:figure4}(b), the relative density depletion up to -8.44\% on 12$^{th}$ of May is only observed with respect to the red line. However, it is clear that the poststorm density in the afternoon local time decreased to as low as the cyan horizontal line, despite an increase in solar EUV flux throughout the event (Figure \ref{fig:figure3}).}

\textcolor{black}{Figure \ref{fig:figure4}(c \& d) show the density variation below -45 $^\circ$N in morning and evening local time. A seasonal variation in density can be clearly seen in the figures as expected (with larger density in summer or northern hemisphere). The effect of solar EUV flux is not as prominent in southern polar region since it was winter polar region during the event. In southern polar regions, the poststorm density decreased nearly to prestorm values despite an increase in solar EUV flux throughout the event, but no relative depletion \textit{or} ``overcooling'' is observed on 12$^{th}$ of May. The relative density depletion of about -5\% is also observed between 0-45 $^\circ$N considering only the red line as referenced value (Figure not shown).}

\begin{figure*}[hbt!]
\includegraphics[width=5.5in,height=3.8in,trim=1.2cm 0.3cm 0.3cm 1.5cm, clip]{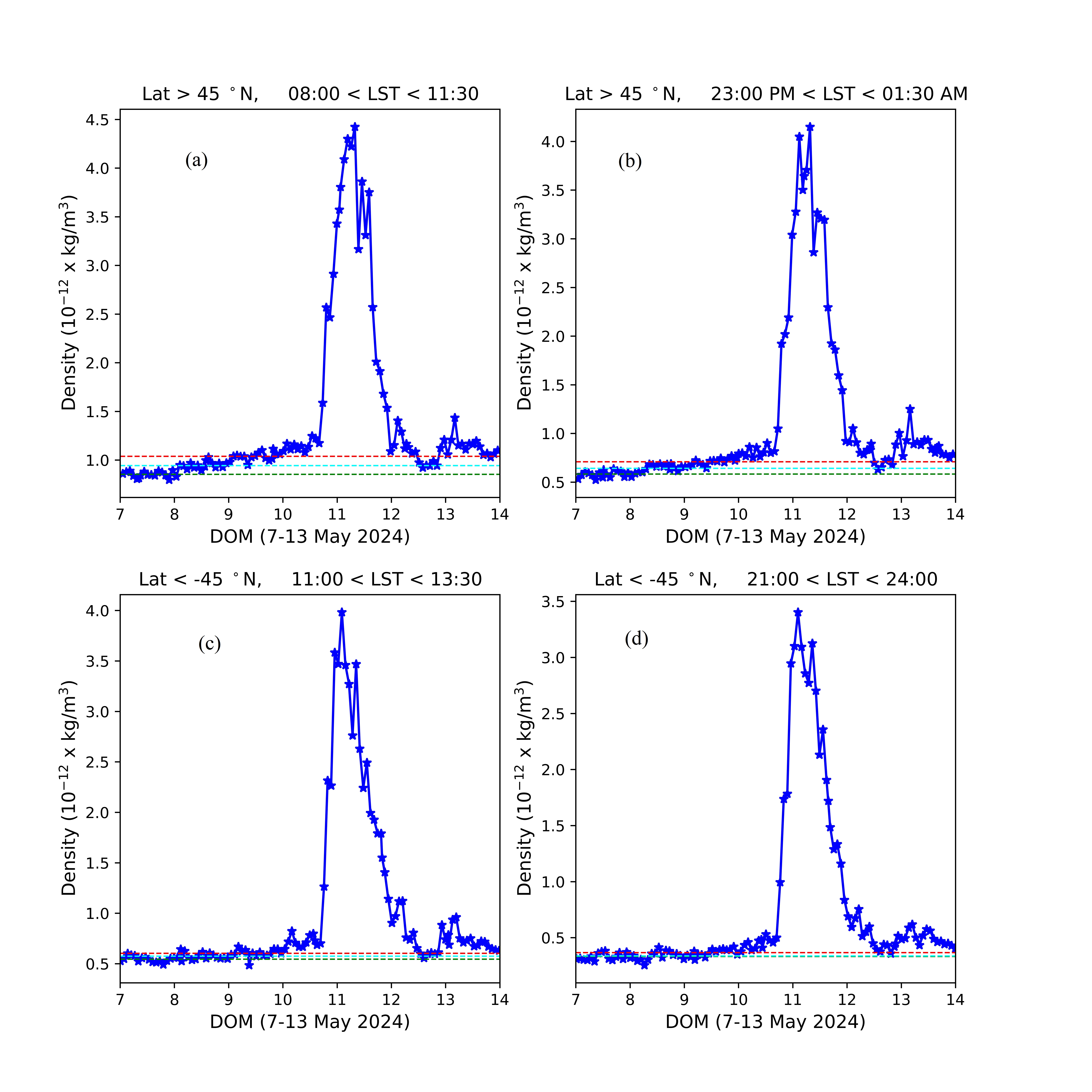}
\centering
\caption{Thermospheric density perturbations as estimated by Swarm-B (a \& b) northern polar region, and (c \& d) southern polar region, from 7-13 May 2024. The green, cyan, and red horizontal dotted lines represent the daily mean density values for 7 May, 8 May, and 9 May, respectively.}
\label{fig:figure5}
\end{figure*}

\begin{figure*}[hbt!]
\includegraphics[width=5.5in,height=3.8in,trim=1.2cm 0.3cm 0.3cm 1.5cm, clip]{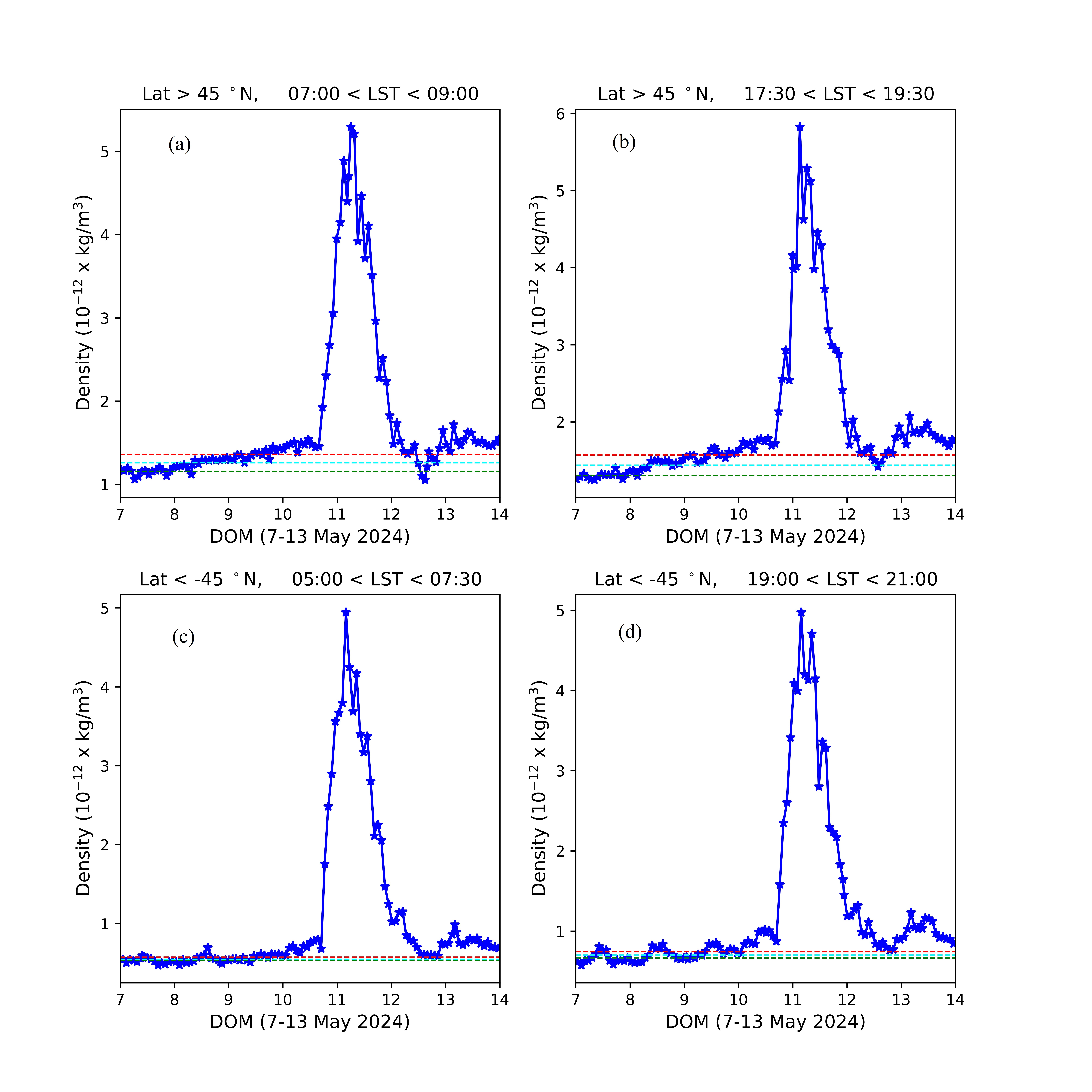}
\centering
\caption{Thermospheric density perturbations as estimated by Swarm-C (a \& b) northern polar region, and (c \& d) southern polar region, from 7-13 May 2024. The green, cyan, and red horizontal dotted lines represent the daily mean density values for 7 May, 8 May, and 9 May, respectively.}
\label{fig:figure6}
\end{figure*}

\justifying
\textcolor{black}{The thermospheric density variations using Swarm-B, and Swarm-C observations are also shown using similar approach in Figure \ref{fig:figure5} and Figure \ref{fig:figure6}, respectively. The same calculations are applied for Swarm-B and Swarm-C to calculate the relative density depletion as for Swarm-A. It can be seen in Figure \ref{fig:figure4} and Figure \ref{fig:figure6} that, density perturbations in Swarm-A and Swarm-C are almost identical. Swarm-C observed local early morning relative depletion near the same UT hours of -9.1\%, -16.67\%, and -23\% in \textcolor{black}{northern} polar region considering the green, cyan, and red line as reference values, respectively (Figure \ref{fig:figure6}(a)). Figure \ref{fig:figure5}(a-d) shows that Swarm-B also observed relative depletion in thermospheric density near 515 km altitude which reached up to -11.37\% in northern polar region near 14 UT on 12$^{th}$ of May between 08:00-11:30 LST with respect to red horizonal dotted line. In the nightside (23:00 PM-01:30 AM LST), Swarm-B observed a relative depletion up to -11.15\%  considering the red line as reference value near the same UT in northern polar regions. Both Swarm-B, and Swarm-C did not observe any ``overcooling'' in the winter or southern hemisphere during the event, however, the poststorm density decreased as low as the prestorm values on the 12$^{th}$ of May.}

As mentioned earlier, the storm time thermospheric NO radiative cooling plays a very important role in balancing the enhanced kinetic energy and modulating density of thermosphere. Figure \ref{fig:figure1}(d) shows that NO IRF increased up to 8-10 times during the storm time (11$^{th}$ of May) in comparison to the pre-storm (8 and 9 May) periods. It is also important to notice that, the enhanced thermospheric density during this superstorm almost came back to its prestorm values between 2-4 UT in norther polar regions on 12$^{th}$ of May (Figure \ref{fig:figure2}). However, the thermospheric NO radiative cooling (NO IRF) remains at an enhanced value of 2-4 times for the entire day on 12$^{th}$ of May in comparison to its pre-storm value. \textcolor{black}{This large amount of NO radiative cooling in thermosphere is potentially responsible for the fast recovery of post-storm thermospheric density.} This unique thermospheric overcooling by NO on May 12 in the recovery phase of the storm could be the primary reason for the observed post-storm depletion in thermospheric density below the pre-storm values. \textcolor{black}{In southern hemisphere, the density has decreased to prestorm values despite an increase in solar EUV flux throughout the event but no overcooling effect is observed. This hemispherical variance aspect needs to be further investigated in the future.} \citeA{lei2012overcooling} reported the overcooling of thermosphere for the first time during two most severe geomagnetic storms of 23$^{rd}$ solar cycle, Halloween storms. It is to be noted that \textcolor{black}{strength and the duration of the Halloween storms is comparable to the May 2024 event making it a primary candidate to witness the overcooling of thermosphere \cite{ranjan2023aspects,bag2024thermospherica}.} A comparison of thermopsheric NO overcooling between the Halloween storms and the May 2024 superstorm has also been presented in the next section.

\subsection{Thermospheric NO overcooling during Halloween storm and its comparison with May 2024 superstorm}

\begin{figure*}[hbt!]
\includegraphics[width=5.5in,height=5.5in,trim=0.2cm 1cm 0.2cm 2cm, clip]{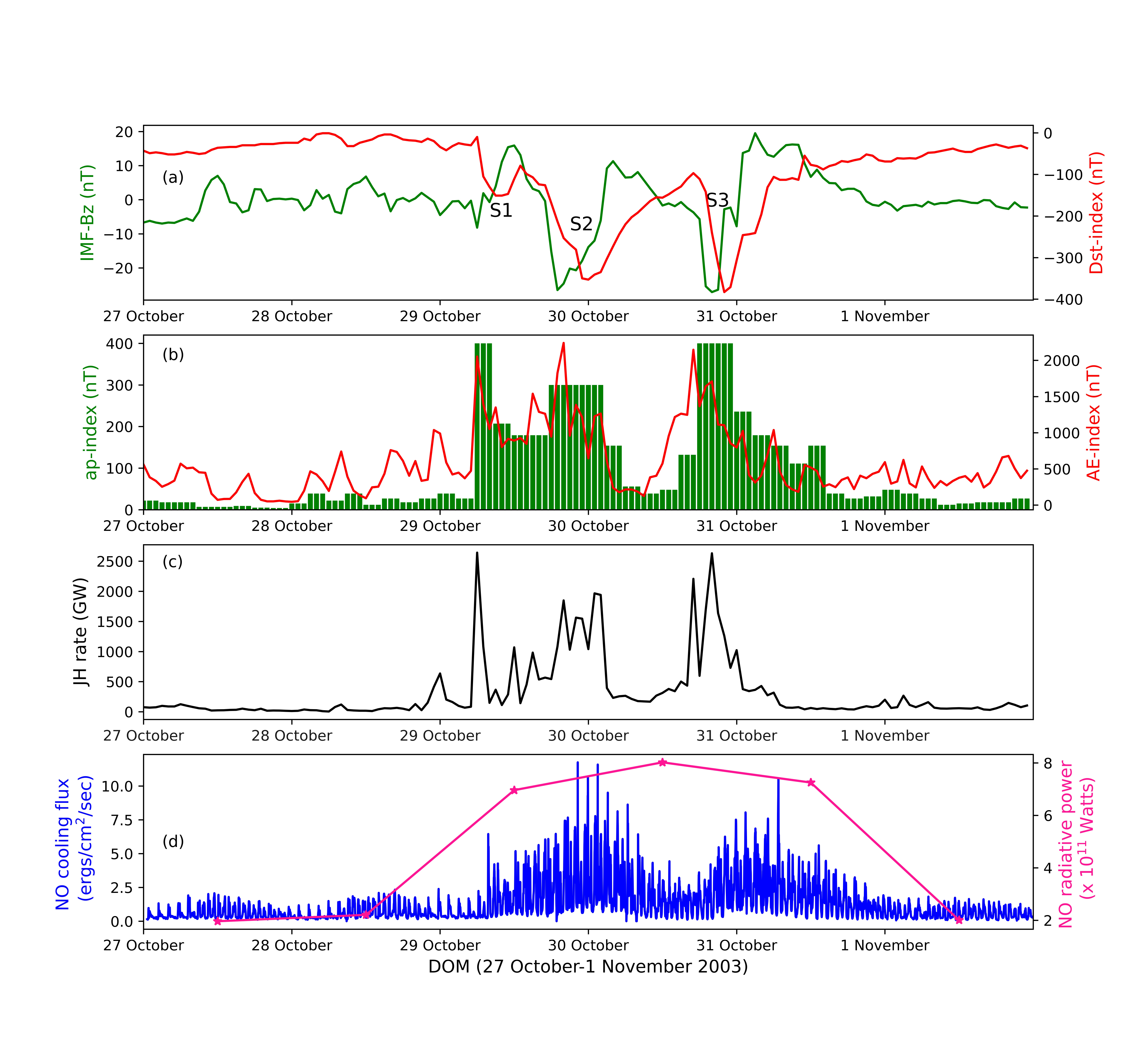}
\centering
\caption{Variations in (a) North-south component of interplanetary magnetic field (IMF-B$_z$) (green), and Dst-index (red), (b) ap-index (green) and AE-index (red), (c) estimated Joule heating rate in northern winter hemisphere, and (d) TIMED/SABER observed NO infrared radiative flux or NO IRF (blue) and associated daily radiative power (magneta) throughout the Halloween storms (27 October-1 November, 2003).}
\label{fig:figure7}
\end{figure*}

\justifying
The solar wind and geomagnetic conditions during the Halloween storms are shown in Figure \ref{fig:figure7} (a-b). The space weather aspects of these storms have been studied extensively in the past \cite{tsurutani2005october,mannucci2005dayside,bergeot2011impact,pedatella2016impact,simpson2020estimating,codrescu2022storm}. Halloween storms were comprised with one intense (S1), and two severe (S2 and S3) geomagnetic storms (Figure \ref{fig:figure7}(a)). Thermospheric NO radiative cooling during the Halloween storms have also been studies recently \cite{ranjan2023aspects,bag2024thermospherica}. It was observed that, besides the large intensity of the S3, the NO radiative cooling were more enhanced during S2 due to its large main phase duration. Figure \ref{fig:figure7}(c) shows the estimated Joule heating rates during the Halloween storms by utilizing the 1-hr cadence Dst-index and Polar Cap indices for Northern hemisphere (PCN-index) \cite{knipp2004direct}. Figure \ref{fig:figure7}(d) shows the NO IRF variations during the Halloween storms, which peaks during S2 up to 11.74 ergs/cm$^2$/sec. The NO IRF during S2 and S3 are enhanced 6-8 times larger in comparison to prestorm periods (27 October).

\justifying  
\textcolor{black}{The NO radiative cooling during the recovery phases of S2 and S3 on October 30$^{th}$ and 31$^{st}$ is still 2-4 times higher than prestorm values.} Joule heating rates at the same time (after 6 UT on 31$^{st}$ of October in particular) are comparable to prestorm values. This lead to the overcooling of thermosphere, which resulted in up to -23 to -26 \% of post-storm relative depletion in thermospheric density in the recovery phases of S2 and S3 compared to prestorm values on 27 October \textcolor{black}{\cite{lei2012overcooling,lei2011rapid,chen2018numerical}.} This study seeks to convey the role of thermospheric NO radiative cooling in the thermospheric density perturbations during the May 2024 geomagnetic superstorm. The SABER observed average radiative power by NO radiative cooling during S2, S3, and May 2024 storms are also calculated for comparison. It is found that the average NO radiative power for S2 between 14 UT on 29$^{th}$ of October (main phase starting time for S2) and 18 UT on 30 $^{th}$ of October (main phase starting time for S3) is 8.66 $\times$ 10$^{11}$ Watts. Similarly for S3, the average NO radiative power is 7.66 $\times$ 10$^{11}$ Watts between 18 UT on 30 $^{th}$ October and 0 UT on 1$^{st}$ of November. For the May 2024 geomagnetic superstorm, the average NO radiative power between 17 UT on 10$^{th}$ of May (main phase starting time for the storm; Figure \ref{fig:figure1}) and 12 UT on 12$^{th}$ of May is about 9.27 $\times$ 10$^{11}$ Watts, which is larger in comparison to S2 and S3. The Joule heating rates were not calculated for the may 2024 geomagnetic superstorm because of the unavailability of definitive Dst-index and PCN-index. The peak NO IRF calculated during 11$^{th}$ of May 2024 is also shows an all time high value of 11.84 ergs/cm$^2$/sec in comparison to 11.74 ergs/cm$^2$/sec near late UT hours of 29$^{th}$ of October 2003.

\section{Summary}
In this study, the effect of May 2024 geomagnetic superstorm on the thermospheric NO infrared emissions and thermospheric density is presented. The superstorm which occurred due to a series of powerful and geoeffective CMEs resulted in the precipitation of enormous amount of particle flux and energy into the polar upper atmosphere and also caused a rare post-storm overcooling of thermosphere. The thermospheric NO radiative cooling flux during geomagnetic superstorm was observed at an all time high value of 11.84 ergs/cm$^2$/sec indicating an enhancement of up to 8-10 times in comparison to prestorm quiet period. This huge energy dissipation \textit{or} cooling of thermosphere by the means of infrared radiation resulted in the \textcolor{black}{fast recovery from Joule heating induced enhanced thermospheric density. The NO induced cooling of thermosphere is also active during the recovery phase of the storm, which is likely responsible for the post-storm observed rare thermospheric overcooling and associated relative density depletion of -23\%. The relative depletion in density also depends on the reference day before the superstorm accounting for the variation in solar EUV flux effect. } The thermospheric cooling response during the May 2024 superstorm is compared with the response during Halloween storms. This study brings out key aspects of the rare thermospheric overcooling which are typically elusive during a majority of severe geomagnetic storms. 


\acknowledgments
Authors acknowledge the TIMED/SABER and Swarm-A, -B, and -C science teams for making the datasets publicly available, which have been utilized in this study. We acknowledge the online article about the recent geomagnetic superstorm of May 2024 available on NASA's website (\url{https://science.nasa.gov/science-research/}) and SpaceWeatherLive.com (\url{https://www.spaceweatherlive.com/}). The authors also acknowledge the PyNRLMSISE-00 (Python interface for the NRLMSISE-00 empirical neutral atmosphere model) available on GitHub (\url{https://github.com/st-bender/pynrlmsise00}), which also has been used in this study.

\section*{Open Research Section}
The authors wish to express their sincere thanks to TIMED/SABER (\url{https://saber.gats-inc.com/browse_data.php}) \cite{esplin2023sounding} and Swarm (\url{ftp://thermosphere.tudelft.nl/}) \cite{siemes2023new} observed datasets utilized in this study. We also acknowledge the solar wind parameters and geomagnetic indices datasets used in this study, which are available at \url{https://omniweb.gsfc.nasa.gov/form/dx1.html}, and \url{https://isgi.unistra.fr/indices_asy.php}. The authors also acknowledge the PyNRLMSISE-00 (Python interface for the NRLMSISE-00 empirical neutral atmosphere model) available on GitHub (\url{https://github.com/st-bender/pynrlmsise00}), which also has been utilized in this study.



%
%

\bibliography{agusample.bib}

%
%
%
%
%

\end{document}